\documentclass[preprint,showkeys,preprintnumbers,amsmath,amssymb]{revtex4}

\usepackage{graphicx}
\usepackage{dcolumn}
\usepackage{bm}

\begin{document}
\title{Minimum entropy production closure of the photo-hydrodynamic equations for radiative heat
transfer}
\author{Thomas Christen}
 \email{thomas.christen@ch.abb.com}
\author{Frank Kassubek}
\affiliation{ABB Schweiz AG, Corporate Research, \\
Im Segelhof, CH-5405 Baden-D\"attwil, Switzerland}


\begin{abstract}
In the framework of a two-moment photo-hydrodynamic modelling of
radiation transport, we introduce a concept for
the determination of effective radiation transport coefficients
based on the  minimization of the local entropy
production rate of radiation and matter. The
method provides the nonequilibrium photon distribution from which the
effective absorption coefficients and the variable Eddington factor
(VEF) can be calculated.  The photon distribution depends on the
frequency dependence of the absorption coefficient, in contrast to the
distribution obtained by methods based on entropy maximization. The
calculated mean absorption coefficients are not only correct in the
limit of optically thick and thin media, but even provide a reasonable
interpolation in the cross-over regime between these limits, notably
without introducing any fit parameter.  The method is illustrated and
discussed for grey matter and for a simple example of non-grey matter
with a two-band absorption spectrum. The method is also briefly
compared with the maximum entropy concept.

\end{abstract}

\maketitle

\section{Introduction}
\label{Intro}
Excessive effort is required for modelling and simulation of radiation
heat transfer in media with complex optical absorption spectra, like hot gases or plasma
\cite{Tien1968}. If the radiation model is part of a
larger model for hot, compressible mixtures of various chemically reacting
species consisting of complex ions, electrons, neutral molecules etc.,
and subject to transonic and turbulent flow, an exact treatment is not nearly possible.
Applications range from arc physics
in welding or electrical switching \cite{Jones1980,Eby1998,Nordborg2008}, atomic explosions,
up to 
astrophysics \cite{Goupil1984}. A frequently used
approximate radiation model is photo-hydrodynamics, which is
based on a two-moment expansion of the radiative transfer equation
and a variable Eddington factor (VEF) closure \cite{Levermore1984}.
This concept can even be realized in a multiband framework, where
the relevant quantities are decomposed according to their spectral properties
\cite{Ripoll2008}.\\ \indent
The main problem of photo-hydrodynamic models is the optimal choice of
the effective transport parameters, i.e., the effective absorption
coefficients and the Eddington factor, which generally depend on
the hydrodynamic and thermodynamic variables. The most prominent
examples are the Planck mean absorption coefficient for optically thin media,
the corresponding Rosseland mean for optically thick media
\cite{SiegelHowell1992}, and the constant Eddington factor of $1/3$ for isotropic
radiation.\\ \indent
Besides these special limit cases, the optimal definition of effective
transport parameters  is not straightforward \cite{Ripoll2001,Ripoll2004}. Accurate
treatment of the general case is particularly important when
 radiation in the cross-over range between optically thin and thick
limit dominates the physical behavior of a system, or if a medium is simultaneously
transparent and opaque for different relevant wavelength bands. A simple
approximate approach to solve that problem can consist in the
construction of fitting expressions that interpolate
between the limit cases. This has been done, for instance, by
Sampson \cite{Sampson1965} for the absorption coefficient and
by Kershaw \cite{Kershaw1976} for the variable Eddington factor.
Another suggestion by Patch \cite{Patch1967} generalizes 
heuristically averages, which are exact for special cases.
A very common approach is based on entropy maximization
\cite{Levermore1984,Minerbo1977,Anile1991}. However,
Struchtrup \cite{StruchtrupUP} has pointed out a weakness of this procedure due
to the neglect of the of radiation-matter coupling for the determination of the
nonequilibrium photon distribution function, which plays indeed a main role for equilibration
\cite{Planck1906}. One of the consequences is that already the Rosseland mean
aborption coefficient in the near-equilibrium case
is not correctly reproduced by a two-moment photo-hydrodynamics
with the maximum entropy closure.
In this paper, we propose to use an entropy {\em production rate} principle, and we
will show that the limit cases are correctly reproduced and
the cross-over between them is provided in a natural
way, i.e., without further model parameters.\\ \indent

Maximization and minimization of the entropy production rate $\dot S$
have turned out to be powerful approaches for modelling many complex
nonequilibrium systems \cite{Martyushev2006}.
We mention that whether the optimum of $\dot S$ is a maximum or a minimum
depends on the type of constraints \cite{Christen2006a} and emphasize that
entropy production optimization is in general not an exact physical law except
near equilibrium, i.e., in linear deviation from equilibrium.
However, it often provides useful
approximate results even far away from equilibrium, provided a
predominance of strongly irreversible equilibration processes. It is also
important for the discussion below that the method is not restricted
to systems in (partial)
local equilibrium, i.e., where the notion of (probably several) local
temperatures can be introduced (e.g., electron and ion temperatures in
a plasma, light pencil temperatures for radiation
\cite{Oxenius1966,Kroll1967}, etc.).  Entropy production
optimization has been shown to be applicable to local
nonequilibrium systems, for instance to Knudsen layers in material
ablation or evaporation processes \cite{FordLee2001,Christen2007a}. In
such cases, the notion of entropy can still be defined \cite{LandauLifshitz}.\\ \indent
Entropy production of radiation has been discussed by Oxenius \cite{Oxenius1966}
and Kr\"oll \cite{Kroll1967}. Various results related to
entropy production principles in radiation have been reported.
Essex \cite{Essex1984} has shown that the entropy production rate is minimum
in a grey atmosphere in local radiative equilibrium. Also this author
has pointed out that a consideration
of the interaction between radiation and matter is crucial because
it contains the appropriate equilibration, i.e.
entropy production, mechanism. 
Later on, W\"urfel and Ruppel \cite{WurfelRuppel1985,Kabelac1994} discussed
entropy production rate maximization by introducing an
effective chemical potential of the photons, related to their interaction with matter.
Finally, we mention Santillan et al.\ \cite{Santillan1998} who showed that for
a constraint of fixed radiation power, the black bodies
are those which maximize the entropy production rate.\\ \indent

This paper is organized as follows. In order to fix notation and to
introduce the relevant quantities, in Sect. \ref{Photohydrodynamics}
the photo-hydrodynamic equations are recalled.
We will consider a system that is characterized by similar
assumptions as in \cite{Patch1967}. Scattering as well as photon time
of flight effects are assumed to be negligible, the matter is
non-relativistic and in thermal equilibrium, and the ordinary index of
refraction is unity. Section \ref{dSdt} provides the basic result,
i.e., Eq. (\ref{maximgen1}) determining the non-equilibrium photon distribution
function. From this distribution function the transport coefficients
can be calculated.  In Sect. \ref{Photo} we show that the approach
provides the Rosseland 
mean in the corresponding limit case.
In Sect.\ \ref{general} cases far from equilibrium are discussed. In particular,
the emission limit, grey matter, and a simple artificial but illustrative example for non-grey matter
are investigated. Our results are compared with the method of entropy maximization in Sect. \ref{Comparison_ME}. 

%
\section{Photo-Hydrodynamics}
\label{Photohydrodynamics}
We start from the Boltzmann equation (which is equivalent to the radiative transfer equation \cite{Tien1968})
for the photon distribution function $f(\vec x,\vec k)$
at location
$\vec x$ and wave number $\vec k = k\vec s$, with direction vector $\vec s $,
$\mid \vec s \mid =1$.
The function $f$ is related to the spectral radiation intensity (radiance) $I$ by
$I= (\omega^{3}/c^{2})\hbar f$,
where $c$ is the velocity of light and $\omega = c k$ is the angular frequency.
Neglecting scattering, the Boltzmann equation for $f$
reads \cite{Tien1968,StruchtrupUP} (here, in contrast to Ref. \cite{Tien1968},
$\kappa (k)$ includes  the material density)
\begin{equation}
\partial _{t} f + c\vec s\cdot \vec \nabla f = - c\kappa (k) (f-f^{(eq)}) \;\;.
\label{BTE}
\end{equation}
Assuming that matter is in thermal equilibrium (at temperature $T$), the emission term contains
the well-known Planck distribution
\begin{equation}
f^{(eq)}(k) = 
\frac{y}{\exp(\hbar c k/k_{B}T)-1} \;\;,
\label{Ffeq}
\end{equation} 
with the photon density of states $y=2/(2\pi)^3$ including two polarization states. (For
a generalization to nonequilibrium matter, e.g., a plasma with
separate electron and ion temperatures, 
the emission term has to be replaced by the appropriate emission source function, and the entropy production of
matter has to be treated in an appropriately generalized way.)\\ \indent
In Eq.\ (\ref{BTE}) scattering is neglected, absorption and spontaneous as well as induced emission
are taken into account in the spectral absorption coefficient
$\kappa (k) = \kappa ^{(0)}(k)[1-\exp (-\hbar c k /k_{B}T)]$.\\ \indent
It is well-known that in local equilibrium particle gas dynamics,
from the Boltzmann transport equation hydrodynamic balance equations for mass, momentum, and energy
balance can be derived.  For photons one can proceed in a similar way but with two
serious restrictions.  First, because the photon number is not
conserved, a continuity equation analogous to mass balance will not
appear. The first moment is related to the energy, which is
proportional to wave number for massless particles.  Secondly, because
the photon gas is generally not near equilibrium, one should in
principle consider a large number of moments of general order and degree
\cite{StruchtrupUP,Struchtrup1998}. Although we will restrict the number of moments to two, the
entropy production method to be used is principally not limited to a specific
number of moments (cf.\ also \cite{StruchtrupWeiss1998}).\\ \indent
In analogy to the P-N model \cite{SiegelHowell1992}, an expansion can
be defined in terms of spherical harmonics,
\begin{equation}
\label{eq:fml}
f_{lm} = \int d^3k\, f(k,\Theta,\varphi) k Y_{lm}^{\ast}(\Theta,\varphi),
\end{equation}
where $\Theta$ and $\varphi$ are the zenith and azimuthal angle of the
vector $\vec{k}$, and the asterisk indicates complex conjugation.
$Y_{lm}$ are the usual spherical harmonic functions with indices
$l$ and $m$. Here, $l$ gives the order of the moment.
Due to the factor $\vec s$ in Eq.\ (\ref{BTE}), the equation of motion
for moments of order $l$ is linked to moments of order $l+1$. We introduce this notation,
because for the analytical calculations below, it is sometimes convenient to use
spherical harmonics.\\ \indent
%
In Cartesian coordinates, the first three moments $f_{lm}$ are associated with energy ($l=0$),
momentum ($l=1$), and radiation pressure tensor ($l=2$).
Energy (photon energy $\hbar \omega = \hbar ck$) and momentum 
(photon momentum $\hbar \vec k$) densities are defined by
\begin{eqnarray}
e (\vec x)   & = & \int d^{3}k \, f\hbar c k \label{edef}\;\;, \\ 
\vec p (\vec x)   & = & \int d^{3}k \, f\hbar \vec s  k  \label{pdef}\;\; .
\end{eqnarray}
The relations of energy and momentum to the moments defined in (\ref{eq:fml}) are, for instance
$e = \hbar c \sqrt{4 \pi} f_{00}$ and $p_{z} = \hbar \sqrt{4 \pi/3} f_{10}$, respectively.
Multiplication of Eq.\ (\ref{BTE}) with $\hbar ck$ and $\hbar k \vec s$
and integration over momentum space gives
\begin{eqnarray}
\partial _{t}e +c^{2}\vec \nabla \cdot \vec p    & = &
P_{e}\label{e}\;\; , \\ 
\partial _{t}\vec p +\vec \nabla  \cdot \Pi  & = & \vec P_{p} \label{p} \;\; ,
\end{eqnarray}
where $\Pi$ is the pressure tensor of the photon gas, having the components
\begin{equation}
\Pi _{mn}  = \int d^{3}k \,s_{m}s_{n} f\hbar c k  \;\;.
\label{Pi}
\end{equation}
The radiation pressure is
$\Sigma _{n=1}^{3}\Pi _{nn}/3 = e/3 $. The source terms of Eqs.\ (\ref{e}) and (\ref{p}), which mediate the interaction of radiation
with matter, are
\begin{eqnarray}
P_{e} & = &  -\hbar c^{2}\int d^{3}k \,  k \kappa (k)
(f-f^{(eq)})\label{Pe}\;\; , \\ 
\vec P_{p} & = & -\hbar c \int d^{3}k\,  \vec k \kappa (k) f \label{Pp} \;\; .
\end{eqnarray}
The radiative heat production $w(\vec x)$, which will appear as a source term
in the energy balance equation for the matter is given by 
\begin{equation}
w = -P_{e} \;\;.
\label{powerprod}
\end{equation}
For a closure of two-moment expansion, the pressure tensor $\Pi$ has to be expressed as a
function of energy $e$ and momentum $\vec p$. Based on tensor symmetry arguments
and on the trace requirement,
one can show that $\Pi$ has the general form \cite{StruchtrupUP}
\begin{equation}
\Pi _{mn}  = e\left(\frac{1-\chi }{2}\delta _{mn} +\frac{3\chi -1}{2} \frac{p_{m}p_{n}}{p^2} \right) \;\;,
\label{Pi2}
\end{equation}
where $\chi (e,p)$ (with $p=\mid \vec p \mid$)
is the so-called variable Eddington factor (VEF) describing radiation anisotropy.
If radiation is isotropic (e.g., in the purely diffusive limit) $\Pi $ must be proportional to the identity matrix, which
implies $\chi = 1/3$. For a beam (e.g., $\vec p $ in $z$-direction),
however, $\chi = 1$ must hold. Therefore,
$\chi $ allows to describe non-diffusive, directed
radiation. The pressure tensor is related to
the second order moments $f_{2 m}$. For instance, assuming $\vec p $ in $z$-direction,
Eq.\ (\ref{Pi}) and $Y_{20}=\sqrt{5/16\pi}(3 \cos ^{2}(\Theta) -1)$ imply the relation
\begin{eqnarray}
\label{Pi2_33}
\Pi_{33}= e \, \chi = \frac{ 1}{3}\left( e  +\sqrt{\frac{16 \pi}{5}}
\hbar c f_{2 0}\right) \;\;,
\end{eqnarray}
which will be needed below. Effective absorption coefficients $\kappa_{e,p}$ can be formally defined by
expressing the source terms (\ref{Pe}) and (\ref{Pp}) by 
%
\begin{eqnarray}
P_{e}   & = & c\kappa _{e} (e^{(eq)}-e)\label{efin}\;\;, \\ 
\vec P_{p}  & = & - c\kappa _{p }\vec p \label{pfin} \;\; .
\end{eqnarray}
%
The effective  absorption coefficients $\kappa _{e,p}(e,p,T)$ generally
depend on $e$, $p$, and $T$. Here, $e^{(eq)}$ is the equilibrium energy density 
\begin{equation}
e^{(eq)} = \frac{4}{c}\sigma T^{4}\;\; ,
\label{eblackbody}
\end{equation}
where $\sigma = 2\pi^{5}k_{B}^{4}/15c^{2}h^{3}\approx 5.67 \,10^{-8} \, $W/m$^{2}$K$^{4}$ is
the Stefan-Boltzmann constant.
The energy radiated by a blackbody per time and area is given by $\sigma T^{4} =
ce^{(eq)}/4$.\\ \indent
%
%
Equations (\ref{e}) and (\ref{p}) define the photo-hydrodynamic equations for the
variables $e$ and $\vec p$ of the photon gas. In order to solve a complete
radiation problem, Eqs.\ (\ref{e}) and (\ref{p}) must be solved together with
the hydrodynamic equations for the matter. This work focuses on the problem of
the determination of the yet unknown effective
transport coefficients $\kappa _{e,p}$ and $\chi $. For this,
the distribution $f(\vec x, \vec k)$ must be known.
Often simple approximate averages are used involving the equilibrium
function $f^{(eq)}$, like the above mentioned Planck mean,
\begin{equation}
\kappa _{\rm Pl}= \frac{\hbar c\int d^{3}k \, \kappa  k f^{(eq)}}{e^{(eq)}}  \;\;,
\label{Pla}
\end{equation}
and Rosseland mean, 
\begin{equation}
\kappa _{\rm Ro} = \frac{ \int d^{3}k \, k^{2} \partial_{k}f^{(eq)}}{\int d^{3}k \, \kappa^{-1} k^{2} \partial_{k} f^{(eq)} }  \;\;.
\label{Ros}
\end{equation}

While the former is a weighted spectral
average of the  absorption coefficient
$\kappa (k)$ that depends on the wave number $k$, the latter is the inverse of a weighted spectral
average of the inverse absorption coefficient $\kappa^{-1}$.
The two cases strongly differ as the
Planck limit is dominated by wave number bands with large $\kappa (k)$
values, while the Rosseland limit is dominated by small $\kappa (k)$. In the associated limit cases radiation is isotropic and the Eddington factor,
being the ratio of pressure and energy density, equals $1/3$.\\ \indent
Prior to the calculation of $f$ and the effective transport coefficients with
entropy production minimization, we mention that the issue
of photo-hydrodynamic boundary conditions for $e$ and $\vec p$ will not be discussed
in this paper. We refer the reader to appropriate literature \cite{SiegelHowell1992,Su2000}.


\section{Minimization of the Entropy Production Rate}
\label{dSdt}
In the following, we determine $f$, and from this $\chi (e,p) $ and $\kappa_{e, p}(e,p) $,
by minimizing the total entropy production rate $\dot S$ under constraints of
fixed values of $e$ and $p$ (we always assume $\vec p$ in $z$-direction).
First, we   derive an expression for
the relevant part of the entropy production rate. The radiation entropy is given by (see, e.g., \cite{StruchtrupUP})
\begin{equation}
S_{\rm rad}(\vec x) = -k_{B} \int d^{3}k \Biggl( f \ln (f/y)-(y+f)\ln (1+f/y)  \Biggr)  \;\;.
\label{S}
\end{equation}
The part of the entropy production rate of the photon gas alone is obtained
by differentiation of Eq.\ (\ref{S}) with respect to time and subsequent replacement of
$\partial _{t}f$ with the help of Eq.\ (\ref{BTE}). This leads to an equation $\partial _{t}S_{\rm rad} + \vec \nabla q_{S_{\rm rad}}=\Sigma$,
with $q_{S_{\rm rad}}$ being the entropy flow density, and the local entropy production rate
\begin{equation}
\Sigma (\vec x) = k_{B} \int d^{3}k \ln \left( y/f+1\right)c\kappa (f^{(eq)}-f)  \;\;.
\label{sdot2}
\end{equation}
The total matter-radiation
system includes the hydrodynamic equations for the matter. The entropy production rate of matter
contains a matter-specific radiation-independent part, which is constant under variation of $f$ and is
thus not of interest here. Additionally, the energy balance of matter contains the power density
Eq.\ (\ref{powerprod}), which can be associated with a local entropy production rate, $w/T$, where $w$ is
obtained from Eq.\ (\ref{Pe}).
The $f$-dependent part of the total local entropy production rate of the radiation-matter
system is thus $\dot S= \Sigma +w/T$. 
Using Eq.\ (\ref{Ffeq}) for replacement of the temperature $T$ leads finally to (see, e.g., \cite{StruchtrupUP})
\begin{equation}
\dot S (\vec x) = -k_{B} c \int d^{3}k \ln \left( \frac{y/f+1}{y/f^{(eq)}+1} \right)\kappa (f-f^{(eq)})  \;\;.
\label{Sdot}
\end{equation}
Optimization of $\dot S$ subject to the constraints of fixed 
$\vec p$ and $e$ implies that the variation of
\begin{equation}
\dot S - \frac{ k_B}{\hbar} \lambda_{e} \Biggl( e (\vec x)   -  \int
d^{3} k \, f \hbar c k \Biggr) - \frac{ k_B c}{\hbar}\vec \lambda_{p} \cdot
\Biggl( \vec p (\vec x)   - \int d^{3}k \, f\hbar \vec k \Biggr) 
\label{sdottot}
\end{equation}
with respect to $f$ vanishes. It follows from the form of Eq. (\ref{Sdot})
that the optimum is a minimum, which we confirmed also numerically.
In Eq.\ (\ref{sdottot}) dimensionless
Lagrange parameters $\lambda _{e}$ and $\vec \lambda _{p}$ have been
introduced. Variation gives
\begin{equation}
y\frac{f-f^{(eq)}}{(y+f)f}-\ln \left(\frac{y/f+1}{y/f^{(eq)}+1}
\right) +\lambda_{e} \frac{ k}{\kappa (k)}
+\vec \lambda _{p} \cdot \frac{  \vec k}{\kappa (k)} =0 \;\;.
\label{maximgen1}
\end{equation}
This is the central result of this paper.  The appearance of $\kappa (k) $ expresses
radiation-matter interaction as the entropy generation process.\\ \indent
The radiation heat transfer model
proposed in this paper can now be summarized. Equation (\ref{maximgen1})
delivers implicitly $f(k,T,\lambda _{e},\vec \lambda_{p})$. Using this $f$, from the constraints (\ref{edef}) and (\ref{pdef})
one can derive $e(\lambda _{e},\vec \lambda_{p},T)$ and $\vec p (\lambda _{e},\vec \lambda_{p},T)$, as well as
$\kappa _{e}(\lambda _{e},\vec \lambda_{p},T)$ and $\kappa _{p} (\lambda _{e},\vec \lambda_{p},T)$, all of them still dependent on
 $\lambda _{e,p}$. These Lagrange parameters are functions of $e, \vec p$ and $T$ by virtue of
$e=e(\lambda _{e},\vec \lambda_{p},T)$ and $\vec p =\vec p (\lambda _{e},\vec \lambda_{p},T)$. Appropriate replacement leads then
to the wanted functions $\chi (e,p,T)$ and $\kappa _{e,p}(e,p,T)$.\\ \indent
Although this procedure can be cumbersome, for a given material (gas, plasma, etc.) it can be done once, and the
results can then be stored in look-up tables or
described by appropriate fit functions, which can be used further in the simulation of the hydrodynamic approximation.
In the following section we discuss specific cases that are simple enough for an analytical treatment.\\ \indent
%
%
Before we discuss the results, we anticipate a remark on the maximum
entropy approach (see e.g.\ \cite{Ripoll2001,StruchtrupUP}), where $S_{\rm rad}$ takes then the part
of $\dot S$ in Eq.\ (\ref{sdottot}), which has to be optimized. Because, in contrast to $\dot
S$ in Eq. (\ref{Sdot}) the entropy $S_{\rm rad}$ in Eq. (\ref{S}) does
not depend on the absorption spectrum $\kappa (k)$, the resulting ME
distribution function $f_ {\rm ME}$ does neither. Hence, in
contrast to our approach, the ME approach does not take into
account radiation-matter interaction explicitly at the level of the photon
distribution function. 
\\ \indent
As a side remark, we mention that irreversibility, due to radiation-matter
coupling, again enters in both approaches at the hydrodynamic level,
i.e., when Eqs. (\ref{e}) and (\ref{p}) are solved together with hydrodynamic equations for
the matter. Therefore, eventually both methods exhibit irreversibility.
But ME assumes that radiation is in a conditional maximum entropy state, while our approach
goes one step further by considering nonequilibrium states away from the entropy maximum.   
%
\section{Radiation near thermodynamic equilibrium: Rosseland limit}
\label{Photo}
If the deviation from equilibrium can be approximated by linearization
('near equilibrium' or 'weak nonequilibrium' of the radiation),
one can expand Eq.\ (\ref{maximgen1}) in a Taylor series
with respect to $\delta f =f-f^{(eq)}$ and keep only the leading order terms.
In this section we will use spherical moments
(\ref{eq:fml}), in order to show that the approach yields the Rosseland limit
up to arbitrary order $l$ of moments. Furthermore, we make an expansion to second
order in $\delta f$, which yields for the VEF the first nontrivial order in $p$
beyond the constant value of $1/3$.\\ \indent
Vanishing variation of entropy production with constraints of
fixed moments $f_{lm}$ reads
\begin{equation}
\delta \left[ \dot S 
- \sum_{l,m} k_B c \lambda_{l m} \left(f_{l m} - \int d^3k\, k f Y_{l m}^{\ast} \right)
\right] = 0\;\; .
\end{equation}
Here, $\delta$ denotes the variation with respect to $f$ and 
$\lambda_{lm}$, the constant factor $k_B c$ in the
constraints is used to obtain dimensionless Lagrange-multipliers $\lambda _{lm}$ with indices $l$ and $m$,
and the sum extends over all
moments that are to be constrained. By expanding the expression in $\delta f$ up to second
order and solving in second order of the Lagrange multipliers, we obtain
\begin{equation}
\label{eq:deltaf} 
\delta f = - f^{(eq)}(y+f^{(eq)})\; \mu 
 +  3 (y+2 f^{(eq)}) f^{(eq)}(y+f^{(eq)}) \; \frac{\mu ^{2}}{4} \;\; 
\end{equation}
with $\mu = (k/2y\kappa (k))\sum \lambda_{l m}  Y_{l m}^{\ast} $, which must be real.
The unknown Lagrange multipliers $\lambda_{l m}$ must be determined
from the constraints $f_{l m}=\int d^3k\, k (f^{(eq)}+\delta f)
Y_{l m}^{\ast}$. Insertion of $\delta f$ from Eq.\ (\ref{eq:deltaf}) gives
\begin{equation}
\label{eq:flm_of_lambda}
f_{l m} = \delta_{m0} \delta_{l0} f^{(eq)}_{00} + C_1  \lambda_{l m}^{\ast } 
+   C_2 \int
d\Omega\, Y_{l m}^{\ast } \left(\sum_{lm} \lambda_{l m} Y_{l m}^{\ast} \right)^2\;\; ,
\end{equation}
with $f_{00}^{(eq)}=e^{(eq)}/\hbar c \sqrt{4 \pi}$ and 
\begin{eqnarray} 
C_1 &=& \frac{1}{2}\frac{k_B T}{\hbar c}\int_0^\infty dk\, k^4 \frac{ \partial_k f^{(eq)}}{ \kappa}\;\;, \label{C1}\\
C_2 &=& \frac{3}{16}\left(\frac{k_B T}{\hbar c}\right)^{2}\int_0^\infty dk\, k^5 \frac{
   \partial^{2}_{k} f^{(eq)}}{\kappa^2} \;\;. \label{C2}
\end{eqnarray}
We used $f^{(eq)}(y+f^{(eq)})\hbar c = -yk_{B}T\partial_{k}f^{(eq)}$, which simplifies both terms
on the right hand side of Eq.\ (\ref{eq:deltaf}). 
Solving Eq.\ (\ref{eq:flm_of_lambda}) for the Lagrange multiplier
$\lambda_{lm}^*$ and inserting into Eq.\ (\ref{eq:deltaf}), gives
to leading order in $\lambda$
%
\begin{equation}
\label{eq:delta_f_Rosseland}
\delta f = \frac{ k_B T \partial_k f^{(eq)}}{2\hbar c} \frac{ k}{\kappa(k)} \sum_{lm}
\frac{ f_{l m}^*-\delta_{m0} \delta_{l0} f^{(eq)}_{00}}{C_1} Y_{l m}^* \;\;.
\end{equation} 
Generalizing (\ref{efin}),(\ref{pfin}), effective absorption coefficients
for mode $l$ and $m$ can be defined by
\begin{equation}
\label{Plm}
P_{lm}=-\kappa_{lm} \left(
f_{lm} - \delta_{m0} \delta_{l0}
\right)
\;\; .
\end{equation}
Inserting (\ref{eq:delta_f_Rosseland}) into the source term
$P_{lm}=-\hbar c\int d^3k\, \vec k \kappa f$ [see Eqs. (\ref{Pe}), (\ref{Pp})], we
obtain
\begin{equation}
\kappa_{ l m} = 
\frac{ \int_0^\infty dk\, k^4  \partial_k f^{(eq)} }{
  \int_0^\infty dk\, k^4  \kappa^{-1}  \partial_k f^{(eq)}} =
\kappa_{\rm Ro} \;\; .
\end{equation}
%
%
%
In this approximation, the absorption coefficients do not depend on the order of the moment and
are equal to the Rosseland mean
(\ref{Ros}). Hence, $\kappa _{e} = \kappa _{p} = \kappa _{\rm Ro}$ near equilibrium.
This is not very astonishing, because the Rosseland average {\em is} the appropriate
effective radiation absorption coefficient near local thermal
equilibrium and, in this limit, the entropy production rate principle
holds exactly.\\ \indent

Let us now calculate the Eddington factor $\chi $ near equilibrium. 
Only constraints with $l \leq 1$ have to be used. Aligning the
$z-$axis of the coordinate system along the momentum $\vec p$,
the Eddington factor is given by $\chi = \Pi _{33}/e$, and all
moments with $m\neq 0$ vanish.
We will thus ignore the index $m$ in the following (e.g.,
$f_0 \equiv f_{00}$, $f_1 \equiv f_{10}$, etc.).
The two equations
(\ref{eq:flm_of_lambda}) for $l=0,1$ can be solved for
$\lambda _{l}$ up to second order in deviation from equilibrium:
\begin{eqnarray}
\label{eq:LagrangeMultSecondOrder}
\lambda_0 &=& \frac{ 1}{C_1} \left(f_0-f_{0}^{(eq)} \right) 
-\frac{ C_2}{2 {C_1}^3} \left[ \left( f_0-f_{0}^{(eq)}+f_1 \right)^2 
+ \left( f_0-f_{0}^{(eq)}-f_1 \right)^2 \right],\\
\lambda _1 &=& \frac{ 1}{C_1} f_1 
-\frac{ C_2}{2 {C_1}^3} \left[ \left( f_0-f_{0}^{(eq)}+f_1 \right)^2 
- \left( f_0-f_{0}^{(eq)}-f_1 \right)^2 \right].
\end{eqnarray}
This allows to calculate from 
Eq.\ (\ref{eq:deltaf}) up to second order
$\delta f$ and $f_{20}$ needed in Eq.\ (\ref{Pi2_33}). One finds $f_{20}=
C_2 f_1^2  /\sqrt{5 \pi} C_1^2$. Using $p_{z}\equiv p = c \sqrt{4 \pi/3} f_{1}$, the Eddington factor
becomes up to quadratic order away from equilibrium
\begin{equation}
\label{eq:chi_final}
\chi \approx  \frac{ 1}{3} +\left(\frac{k_B T}{\hbar c}\right)^{4}\frac{C_2}{75 C_1^2}
\left( \frac{c p}{e^{(eq)}} \right)^{2}\;\; .
\end{equation}
Thus, near equilibrium, the  Eddington factor is $1/3$ as it must be, and the VEF
grows quadratically with $p$. The curvature depends on $\kappa (k)$
unless the medium is grey, as will be discussed in Sect.~\ref{greymatter}.\\ \indent
%

\section{General nonequilibrium radiation}
\label{general}
\subsection{The emission limit}
If the photon density is so small that absorption can be neglected, the emission approximation
can be applied. Formally, this refers to the limit $f \ll f^{(eq)}$, or $e/e^{(eq)}\to 0$,
such that the source term on the right
hand side of Eq.\ (\ref{BTE}) becomes $c\kappa (k)f^{(eq)}$. It is clear from Eq.\ (\ref{Pe}) that the matter is then
cooled, with a cooling power density
$w = - c\kappa _{\rm Pl} e^{(eq)}$ that contains the Planck mean (\ref{Pla}) for $\kappa _{e}$.
In order to determine $\kappa _{p}$, we must go to first order in $f$ beyond the Planck limit.
Expansion of Eq.\ (\ref{maximgen1}) up to leading order of $f/f^{(eq)}$ gives
\begin{equation}
f = \frac{\kappa (k)}{\lambda _{e} k+ \vec \lambda_{p} \cdot \vec k }\; f^{(eq)}(k)\;\;.
\label{fPlanck}
\end{equation}
Substitution of $f$ in Eqs.\ (\ref{edef}) and (\ref{pdef}) leads to
implicit equations for $e(\lambda _{e},\vec  \lambda_{p})$ and
$p(\lambda _{e},\vec  \lambda_{p})$. 
They  can be solved in leading order of $\lambda_{p}$
corresponding to the case of small momentum $\vec p$ or almost
isotropic radiation.
One finds $\lambda _{e} = \hbar c \tilde \kappa /e$ and $\vec
\lambda_{p} =-3\hbar c^{2} \tilde \kappa \vec p / e^{2}$ with $\tilde
\kappa = \int d^{3}k \kappa  f^{(eq)}$. Replacement of $f$ in
Eq.\ (\ref{Pp}) and using these results leads to $\vec P_{p} = -
\widetilde {\kappa ^{2}}c \vec {p}/\tilde {\kappa}$, with $\widetilde
           {\kappa ^{2}} = \int d^{3}k \kappa^{2} f^{(eq)}$. In
           summary:
\begin{eqnarray}
\kappa _{e} & = & \kappa _{\rm Pl} \label{emissionke} \\
\kappa _{p} & = & \frac{\widetilde {\kappa ^{2}}}{\tilde {\kappa }} \label{emissionkp} \;\; .
\end{eqnarray}
Energy and momentum relaxation have different effective relaxation constants in the limit of small $e$
for non-grey matter. From Eq.\ (\ref{fPlanck}) one obtains with Eq.\ (\ref{Pi})
the Eddington factor $\chi = 1/3$ in this limit, i.e., in the emission approximation radiation is also isotropic.
%

\subsection{Grey matter}
\label{greymatter}
Grey matter is characterized by wave number independent spectral
absorption, $\kappa (k)\equiv$ const.
The effective absorption coefficients are $\kappa _{e} \equiv \kappa _{p}\equiv \kappa$,
and the nonequilibrium distribution $f$ does not depend explicitly on the $\kappa$ value. 
This follows from the proportionality of the entropy production rate to $\kappa $ in Eq.\ (\ref{maximgen1}).
However, note that
in the framework of the hydrodynamic equations (\ref{e}) and (\ref{p}), $f$ depends {\em implicitly}
on the $\kappa $ value via its $e$ and $p$ dependence. \\ \indent

\begin{figure}[h]
   \centering
    \rotatebox{0}{
        \resizebox{0.8\textwidth}{!}{
            \includegraphics{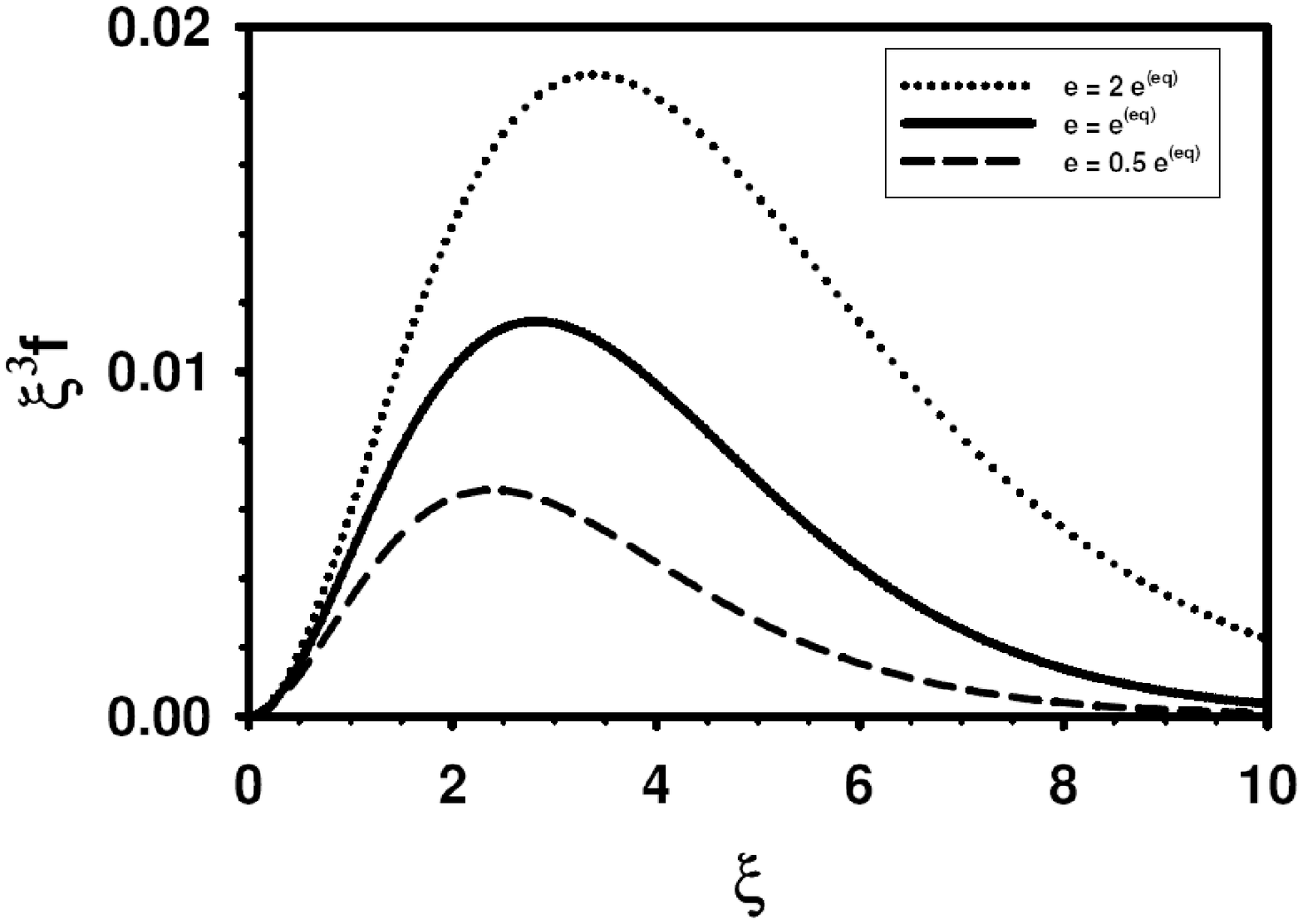}}}
 \caption{Function $f \xi ^3$ for $p=0$ as a function
  of $\xi = \hbar c k/k_{B}T$ (in analogy to
  the spectral power density) obtained from the entropy maximization
  method applied to a grey body. $e=e^{(eq)}$ (solid, Planck function),
  $e=2e^{(eq)}$ (dotted), and $e=0.5e^{(eq)}$ (dashed). Larger energy implies a considerably higher photon
  occupation number and only a weak shift to larger wave numbers.}
\label{Fig1}
\end{figure}

\begin{figure}[h]
\centering
    \rotatebox{0}{
        \resizebox{0.8\textwidth}{!}{
            \includegraphics{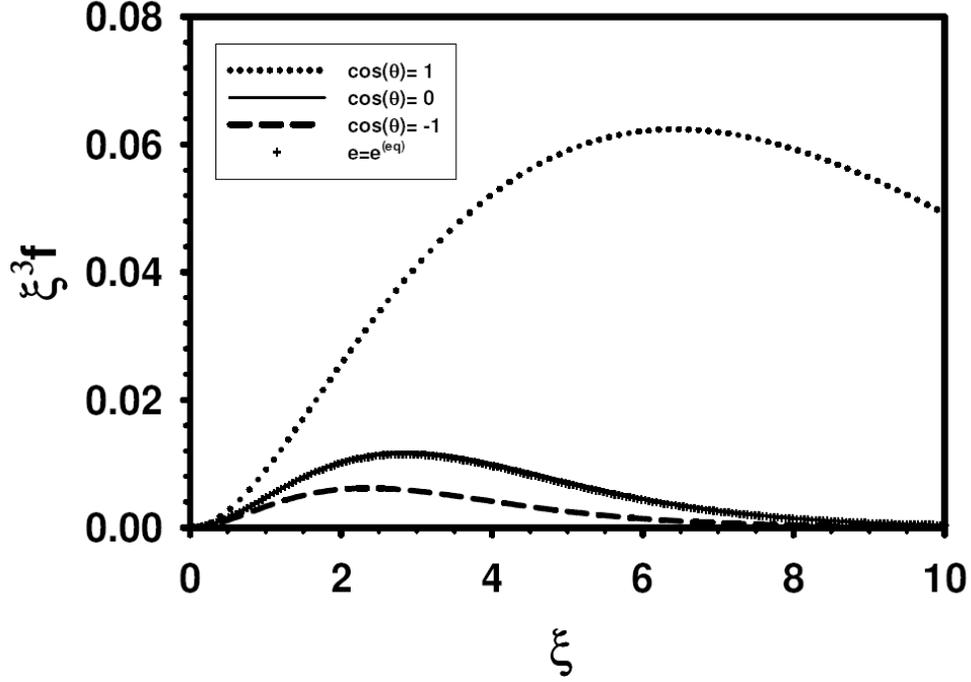}}}
\caption{Function $f \xi ^3$  as function of
 $\xi = \hbar c k/k_{B}T$ and for different values of the cosine of the angle $\Theta$ between $\vec k$ and $\vec p$.
 The value $\cos(\Theta )= 1$ corresponds to photons parallel to $\vec
 p$, while $\cos(\Theta )= -1$ corresponds to
 anti-parallel photons. Here, the energy is roughly equipartitioned between ballistic and diffusive motion, i.e.,
 $e=2 e^{(eq)}$, $p=e^{(eq)}/c$; symbols indicate the equilibrium distribution.}
\label{Fig2}
\end{figure}

We have numerically calculated $f$ for grey matter. In Fig.~\ref{Fig1} a plot of
$\xi ^{3}f $ is shown as a function of $\xi = \hbar ck/ k_{B}T$ for different values of $e$ at $p=0$.
One observes that it is mainly the photon occupation number, which increases for higher energy $e$,
while the shift of the wave number at maximum $f$ towards larger values is comparatively weak.\\ \indent
As the momentum $p$ increases, however the wave number is expected to be more
affected. Figure \ref{Fig2} shows the distribution as a function of wave number for different values of the
direction cosine.
In this example, $pc=e^{(eq)}$ and $e=2e^{(eq)}$, which roughly means that half of the energy is
thermal and half is "ballistic". The function $\xi ^{3}f $  at $\cos (\Theta) =0$ is thus very close
to the equilibrium distribution for $e=e^{(eq)}$, which is also plotted in the figure.
Besides a change in their number density, the photons parallel ($\cos (\Theta) =1$) and anti-parallel
($\cos (\Theta) =-1$) to the beam are shifted towards higher and lower wave numbers, respectively.\\ \indent

\begin{figure}[h]
\centering
    \rotatebox{0}{
        \resizebox{0.8\textwidth}{!}{
            \includegraphics{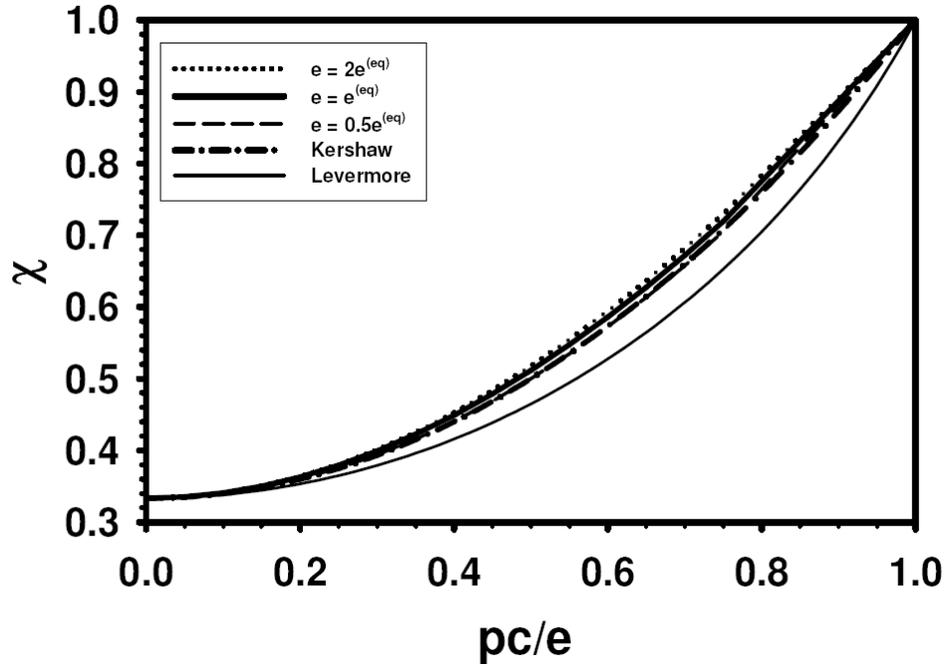}}}
\caption{VEF of a grey body for the different cases listed in the legend.
Deviations from $\chi _{K}$ are below the percentage range, deviations from the Levermore-Eddington factor
$\chi _{ME}$ are larger.}
\label{Fig3}
\end{figure}

The variable Eddington factor (VEF) obtained from our results turns out to be close to the heuristic
proposal by Kershaw \cite{Kershaw1976}, $\chi_{K}= (1+2(cp/e)^2)/3$. In particular, as required
$\chi = 1/3$ for $p=0$ and $\chi \rightarrow 1$ for $pc \rightarrow e$. Figure \ref{Fig3} shows
a plot of $\chi $ as a function of $cp/e$ for various cases.
$\chi _{K}$ seems to be a reasonable approximation for the VEF.
For grey matter the integrals (\ref{C1}) and (\ref{C2}) can
be calculated analytically by partial integration. Equation (\ref{eq:chi_final}) implies then for
the VEF near equilibrium (i.e., $e\approx e^{(eq)}$ and $pc \ll e$) $\chi (e,p)
\approx (1+(3cp/2e)^{2})/3$. \\ \indent

The results obtained from the entropy production method are not only exact in the weak nonequilibrium limit,
but provide the correct VEF in the limit of directed radiation ($p\to e/c$). This supports the conjecture that
the entropy production method, in the present context, can serve as a useful approximation even far from
equilibrium.


\subsection{Non-grey matter}
\label{twoband}
\begin{figure}[h]
\centering
    \rotatebox{0}{
        \resizebox{0.8\textwidth}{!}{
            \includegraphics{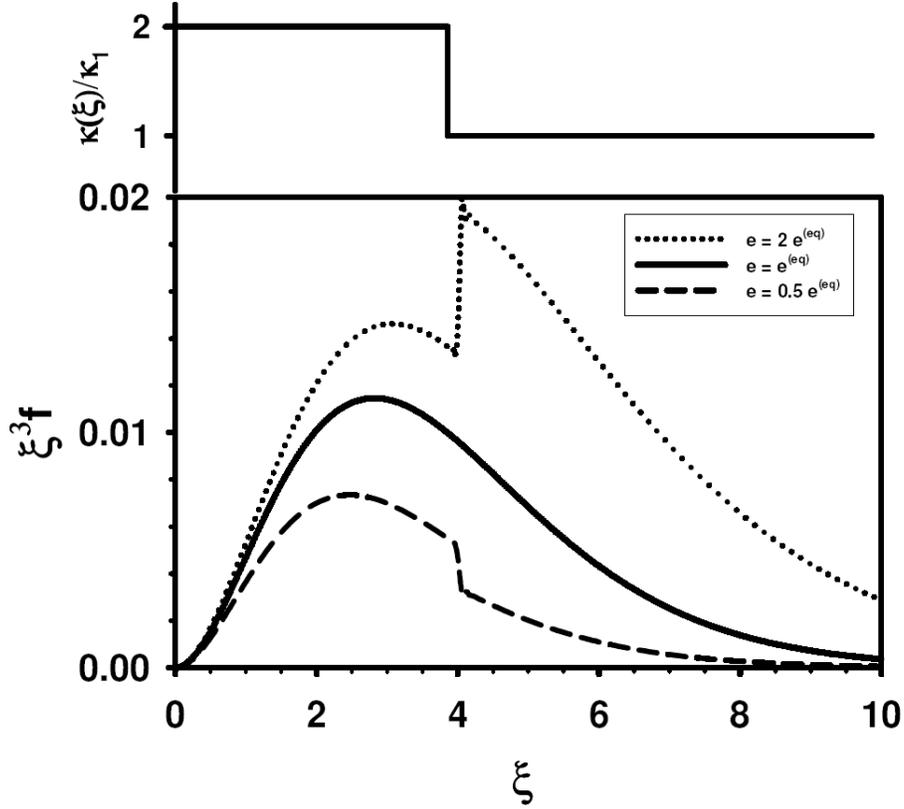}}}
\caption{Top: Step-function absorption coefficient. Bottom: Function $f \xi ^{3}$ for different
 $e/e^{(eq)}$-values. The change in $\kappa$ at $\xi = 4$ implies a
  change in the distribution function. Higher $\kappa (k)$ values pulls the distribution function $f$ stronger
  towards the equilibrium distribution $f^{(eq)}$ because of stronger radiation-matter interaction.}
\label{Fig4}
\end{figure}

As a simple example with mainly illustrative purpose, we consider now
an artificial two-band absorption coefficient.
We will use the non-dimensional quantity $\xi $ instead of the wave
number $k$ and consider a step-function absorption-spectrum of the
form $\kappa (\xi ) = 2 \kappa _{1}$ for $\xi < \xi_{c}$ and $\kappa
(\xi) =\kappa _{1}$ for $\xi > \xi _{c}$, where $\kappa _{1}$ is
constant (cf.\ Fig.\ \ref{Fig4}).  This artificial spectrum $\kappa
(\xi)$ describes matter-radiation interaction that is stronger for
long than for short wavelengths. To be concrete, $\xi _{c}=4$, which
gives $\kappa _{\rm Pl} = 1.6\,\kappa _{1}$, $\kappa _{\rm Ro} =
1.26\,\kappa _{1}$, and $\tilde {\kappa ^{2}}/\tilde {\kappa } =
1.89\,\kappa _{1}$.\\ \indent One might have expected that for $e\to
0$ (emission limit) the value of $\kappa _{p}$ becomes equal to
$\kappa (\xi \to 0)=2 \kappa _{1}$, which is obviously not the case.
The value $\kappa _{p}=1.89 \,\kappa _{1}$ in this limit can be
understood with the help of Eq.\ (\ref{fPlanck}), which yields
$f\approx f^{(eq)}\kappa (k)e /\tilde \kappa \hbar c k$. Hence $f$ is
a given function of $k$ (or $\xi $), and is proportional to $e$. This
implies that $\kappa _{p}$ is a fixed weighting of the spectral
absorption coefficient.\\ \indent

The function $\xi ^{3}f$ obtained from the minimum entropy production principle
is shown in Fig.~\ref{Fig4} for $p=0$ and different $e/e^{(eq)}$ values.
A similar qualitative dependence of the photon occupation number on $e/e^{(eq)}$
appears as for grey matter. Furthermore, the step in $\kappa (\xi )$ leads to
a step in $f$, such that the nonequilibrium distributions are closer to equilibrium
for $\xi<\xi_{c}$. This pull of the long-wavelength part towards equilibrium can be understood from the larger
$\kappa (\xi)$ value in this region. Indeed, stronger interaction of radiation with matter leads
to stronger equilibration.\\ \indent

\begin{figure}[h]
\centering
    \rotatebox{0}{
        \resizebox{0.8\textwidth}{!}{
            \includegraphics{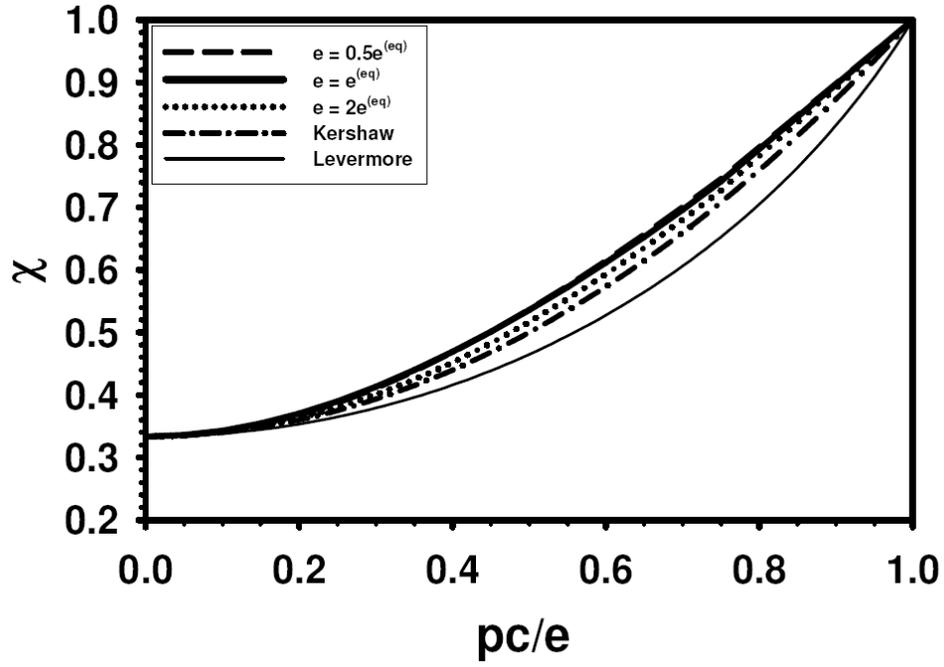}}}
\caption{Same as in Fig.~3, but now for the step-function absorption coefficient of Fig. 4. The deviations
from $\chi _{K}$ are a few percent, while deviations from $\chi _{ME}$ are twice as large.}
\label{Fig5}
\end{figure}

The VEF for this special case of a stepwise absorption spectrum is shown in
Fig.~\ref{Fig5}. The VEF has increased as compared to the grey medium, hence the deviations from Kershaw's VEF
are a little bit larger. The increase of $\chi $ at fixed $cp/e$ might be understood qualitatively as follows,
if one interprets $\chi $ as a measure for the radiation pressure in direction of momentum $\vec p$. Because the
considered absorption spectrum leads to enhanced equilibration of the long wavelength photons, there must be an increased
amount of short wavelength photons contributing to the momentum $\vec p$, which leads to the larger $\chi $\\ \indent

\begin{figure}[h]
\centering
    \rotatebox{0}{
        \resizebox{0.8\textwidth}{!}{
            \includegraphics{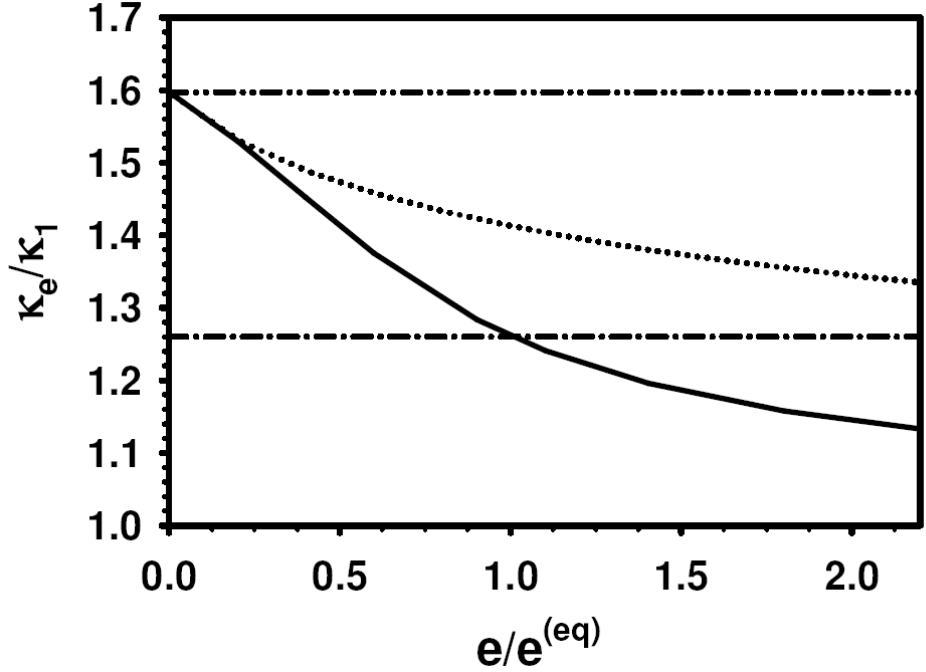}}}
\caption{Effective absorption coefficient $\kappa_e$ (divided by $\kappa _{1}$) at $p=0$ for the step function
  absorption coefficient of Fig. 4. $\kappa_e$ obviously takes the Planck mean value (dashed-double-dotted) at $e=0$
  (emission limit) and the Rosseland mean (dashed-dotted) at
  $e=e^{(eq)}$ (equilibrium radiation).
  For large energies, $\kappa_e$ will converge to $\kappa _{1}$, i.e. to the $\kappa (k)$ value
  at large wave numbers populated by most of the photons. Entropy maximization leads to a mean absorption $\kappa _{\rm ME}$
  (dotted) with the
  correct Planck mean at $e=0$, but with a deviation from the Rosseland mean at $e=e^{(eq)}$.}
\label{Fig6}
\end{figure}

\begin{figure}[h]
\centering
    \rotatebox{0}{
        \resizebox{0.8\textwidth}{!}{
            \includegraphics{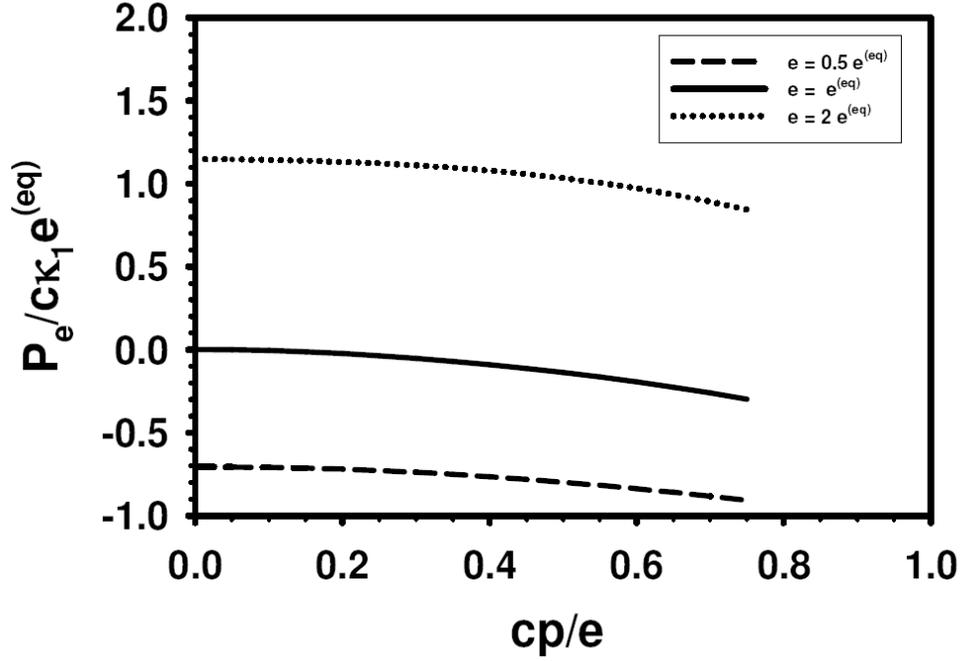}}}
\caption{Energy loss function $P_{e}$ as a function of $cp/e$ for different $e/e^{(eq)}$ values. This figure
illustrates that the zero of $P_{e}(e)$ is located at $e^{(eq)}$ only for equilibrium conditions, i.e., $p=0$. For finite
$p$ the zero slightly shifts, i.e. $P_{e}(e^{(eq)})$ is finite.}
\label{Fig7}
\end{figure}

The normalized effective absorption coefficient $\kappa _{e} / \kappa _{1}$ is shown in Fig.~\ref{Fig6} as a function
of $e/e^{(eq)}$ at $p=0$. In the limits $e\to 0$ and $e=e^{(eq)}$, the Planck and Rosseland mean
are obtained as one expects. For $e\to \infty $ it holds $\kappa _{e}(e,p=0) \to \kappa _{1}$,
as most photons will populate the short wavelength band. For finite $p$, it is inconvenient to discuss $\kappa _{e}(e,p)$
because the zero of $P_{e}(e,p)$ is no longer given by $e = e^{(eq)}$ but shifts as a function of $p$. (Note that
finite $p$ is always associated with nonequilibrium.) An effective $\kappa _{e}$, however, can still be defined as in
Eq.\ (\ref{Pe}), but has a pole at $e^{(eq)}$ and a zero shifted away from $e^{(eq)}$.
This fact is illustrated in Fig.~\ref{Fig7}, which shows that $P_{e}(e,p)$ starts to deviate from
the value at $p=0$ as $p$ increases (particularly the shift of $P_{e}(e^{(eq)},p)$ from zero). Although equivalent,
it might be more convenient to work with $P_{e}$ or with representations as in, e.g., Ref. \cite{Ripoll2008}
than with $\kappa _{e}$, if general simulations of a fully coupled hydrodynamic radiation problem are performed.\\ \indent

\begin{figure}[h]
\centering
    \rotatebox{0}{
        \resizebox{0.8\textwidth}{!}{
            \includegraphics{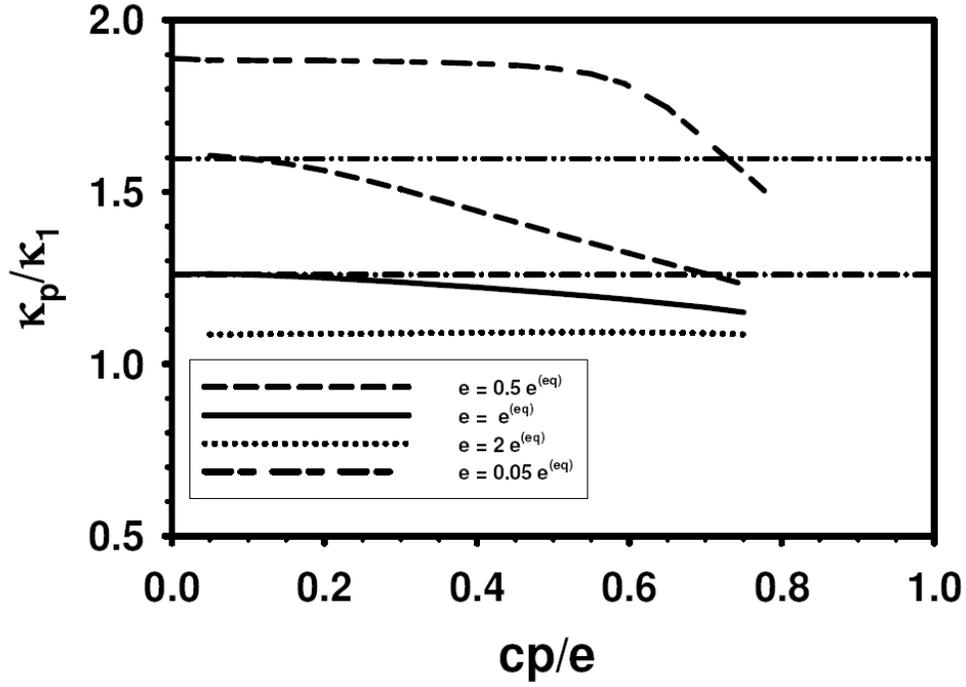}}}
\caption{Effective absorption coefficient $\kappa_p$ (divided by $\kappa _{1}$) as a function of $cp/e$
   for the step function
  absorption coefficient of Fig. 4. The dashed-dotted and the dashed-double-dotted lines indicate the Rosseland and Planck means,
  respectively.}
\label{Fig8}
\end{figure}

Figure \ref{Fig8} shows the effective absorption coefficient $\kappa _{p} (e,p)/ \kappa _{1}$ as a function of $pc/e$ for
different $e/e^{(eq)}$-values. Again, for small $p$, $\kappa _{p}$ is large at small $e$ (emission approximation, where
it goes to the value $1.89$), while
near equilibrium the Rosseland mean is reproduced. At large $e$, $\kappa _{p} $ goes to $\kappa _{1}$
because the distribution function $f$ extends to higher and higher wave numbers.
The same behavior occurs when $cp/e$ becomes large, i.e., $\kappa _{p} $ approaches $\kappa _{1}$,
as can be concluded from the dotted curve in Fig. \ref{Fig2}, which indicates that the distribution function
of directed photons involves large wave numbers.
%
\section{Comparison with the Maximum Entropy Method}
\label{Comparison_ME}
In this section we compare our method with the well-known ME
approach, where the radiation entropy is maximized. Especially,
effective absorption coefficients and VEF are compared for the
different explicit cases treated in the previous sections. 
The basic drawback of ME is the explicit independence of $f$ on $\kappa (k)$.
The implicit dependence on $\kappa $ via radiation-matter coupling at the
{\em hydrodynamic} level (i.e.,via an $e$ and $p$-dependence due to the ME constraints) is not
able to spectrally resolve the absorption properties in the distribution function.
Indeed, one can easily calculate $f$, which is given by \cite{StruchtrupUP}
\begin{equation}
f_{\rm ME} = \frac{y}{\exp{(\hbar c k (\lambda _{e} -\vec \lambda _{p}\vec s))}-1}
\label{fME}
\end{equation}
with $e$ and $p$ dependent Lagrange parameters. At this level, information about the absorption spectrum
is not contained.\\ \indent
On the other hand, according to Eq. (\ref{maximgen1}),
the distribution function $f$ obtained from entropy production minimization does explicitely
depend on $\kappa (k)$.
In subsection \ref{twoband}, it has been shown that there is a significant and easily understandable
effect of wave number dependent $\kappa (k)$ on the distribution function, and thus also on
the effective absorption coefficients and on the VEF.\\ \indent  
First, we note that because of the $\kappa $-independece of $f_{\rm ME}$,
the associated VEF is a unique function of $e$ and $p$,
\begin{equation}
\chi _{\rm ME} = \frac{5}{3}-\frac{4}{3}\sqrt{1-\frac{3}{4}\frac{c^{2}p^{2}}{e^{2}}} \;\; ,
\label{Levermore}
\end{equation}
which is the well-known Levermore-Eddington factor \cite{StruchtrupUP}.
It is also plotted in Figs. \ref{Fig3} and \ref{Fig5} for
comparison. For grey matter the Eddington factors obtained by our approach and by ME are different, but probably similar
enough to justify the use of $\chi _{\rm ME}$. For non-grey matter the deviations increase.\\ \indent
The difference in the effective
absorption coefficients $\kappa _{e,p}$ is more important.  In the limit case of weak
nonequilibrium, where entropy production minimization reproduces the
correct Rosseland mean, ME gives (i.e., up to leading order in $f_{\rm ME}-f^{(eq)}$)
\begin{equation}
\kappa _{\rm ME} = \frac{ \int d^{3}k\, \kappa \, k^{2} \partial_{k}f^{(eq)}}{\int d^{3}k \,  k^{2} \partial_{k} f^{(eq)} }  \;\;,
\label{kappaME}
\end{equation}
which is clearly different from the correct Rosseland mean (\ref{Ros})
(cf. \cite{StruchtrupUP}). Note that the two methods yield different results for the
transport quantities in equilibrium limit, because the deviation from equilibrium, $\delta f = f-f^{(eq)}$,
is different. Although the zero'th order term $f^{(eq)}$ is the same and
$\delta f$ goes to zero, it is $\delta f$ which determines the transport coefficients.
As an example, we have added the curve for $\kappa _{\rm ME} $ in
Fig. \ref{Fig6}. While the Planck limit ($e\to 0$) is correctly
obtained by ME, there is a difference of about $10 \%$ to the
Rosseland limit ($e=e^{eq}$) for our simple example.  
As has been
shown in Ref. \cite{StruchtrupUP}, the ME approach is able to
reproduce the Rosseland mean only if one considers a larger number of
moments, which goes beyond a two-moment photo-hydrodynamic description of the
photon gas with variables $e$ and $ \vec p$.\\ \indent

A remark on the $\kappa $-dependence of our approach is in order. One might conclude from Eq. (\ref{Sdot})
that in the case of vanishing $\kappa $
the minimum entropy production rate approach is not defined, while the maximum entropy approach is
defined because $\kappa $ does not appear.
However, zero $\kappa $ means {\em absence} of equilibration. But maximum entropy requires that equilibration
to the associated maximum entropy state already occurred. Because for zero $\kappa$ there is no process that
drives the photon gas towards the associated maximum entropy state, both approaches encounter a similar
conceptual problem.

\section{Conclusion}
\label{conc}
The description of radiative heat transfer with photo-hydrodynamic
equations requires a closure procedure to restrict the number of coupled
equations. If radiative energy density $e$ and momentum density $\vec p$
are to be considered, effective absorption coefficients and a variable Eddington factor
(VEF) have to be determined as a function of $e$ and $\vec p$.
Using the minimization of the (local) entropy production rate of radiation {\em and}
matter, a closure procedure is introduced that yields the photon distribution function, which
depends on the spectral absorption coefficient, and which
allows for the calculation of transport coefficients in general.\\ \indent
It has been shown that the derived expressions are exact near equilibrium, in the emission
limit, and correctly describe directed radiation with a VEF $\chi = 1$.
Whereas the first fact is not very surprising as
minimum entropy production is known to be exact near equilibrium, the other correct
limiting behavior demonstrates that entropy production optimization can provide sound and
useful results also far from equilibrium.\\ \indent
For the general case, effective
absorption coefficients and a variable Eddington factor are found, which
reasonably interpolate between the limiting cases, notably
without introduction of any fit parameter.
The variable Eddington factor for grey matter is close to the one proposed by Kershaw
\cite{Kershaw1976}, and not far from the Levermore-Eddington factor, which can be derived from
a maximum entropy approach.\\ \indent
In contrast to our approach, the distribution function obtained from the
maximum entropy (ME) method does not contain the absorption spectrum. Hence, ME
cannot describe frequency resolved absorption effects on the distribution function.
An example of such an effect, which is taken into
account by our method, is the enhanced photon equilibration in wave number bands with larger
absorption as discussed in Sect. \ref{twoband}. Another difference is that in the two-moment photo-hydrodynamic
framework, ME is unable to reproduce the Rosseland mean absorption coefficient in the equilibrium limit.\\ \indent
Roughly speaking, the
maximum entropy approach is a zero order approximation in the sense that radiation has equilibrated
to a conditional maximum entropy state. Our approach can be understood as
a first order approximation by assuming that the radiation is away from equilibrium and equilibrates
according the minimum entropy production method.
To summarize, we conjecture that this method may serve as a useful
approach in various radiation heat transfer problems. 
\\ \indent
A benchmarking
and critical discussion of specific application examples will be
necessary in the future. Furthermore, the feasibility of improvements
in the framework of multiband extensions \cite{Ripoll2008} or by
taking into account higher order moments should be investigated. 

\newpage

\end{document}